\documentclass[prb,twocolumn,superscriptaddress,showpacs,floatfix]{revtex4}

\usepackage[dvips]{graphicx}
\usepackage{amsmath}
\usepackage{booktabs}
\usepackage{dcolumn}

\newcommand{\Angrz}{$\text{\AA}^{-1}$}

\newcommand{\tn}{T$_\text{N}$}

\newcommand{\mno}{MnO$_\text{6}$}
\newcommand{\dymo}{DyMnO$_\text{3}$}
\newcommand{\dmo}{DyMnO$_\text{3}$}
\newcommand{\tbmo}{TbMnO$_\text{3}$}

\newcommand{\kf}{k$_\text{f}$}

\newcommand{\vQ}{$\vec{Q}$}

\begin{document}

\advance\vsize by 2 cm

\title{Magnetic order and electromagnon excitations in DyMnO$_3$ studied by neutron scattering experiments}

\author{T. Finger}
\affiliation{II. Physikalisches Institut, Universit\"{a}t zu
K\"{o}ln, Z\"{u}lpicher Str.\ 77, D-50937 K\"{o}ln, Germany}

\author{K. Binder}
\affiliation{II. Physikalisches Institut, Universit\"{a}t zu
K\"{o}ln, Z\"{u}lpicher Str.\ 77, D-50937 K\"{o}ln, Germany}

\author{Y. Sidis}
\affiliation{Laboratoire L\'eon Brillouin, C.E.A./C.N.R.S., F-91191 Gif-sur-Yvette CEDEX, France}

\author{A. Maljuk}
\affiliation{Helmholtz-Zentrum Berlin, Hahn-Meitner Platz 1,
D-14109 Berlin, Germany} \affiliation{Leibniz Institute for Solid
State and Materials Research Dresden, Helmholtzstrasse 20, 01069
Dresden, Germany}

\author{D. N. Argyriou}
\altaffiliation[now at: ]{European Spallation Source, S-22100
Lund, Sweden} \affiliation{Helmholtz-Zentrum Berlin, Hahn-Meitner
Platz 1, D-14109 Berlin, Germany}

\author{M. Braden}
\email{braden@ph2.uni-koeln.de} \affiliation{II. Physikalisches Institut, Universit\"{a}t zu K\"{o}ln, Z\"{u}lpicher
Str.\ 77, D-50937 K\"{o}ln, Germany}

\date{\today}


\begin{abstract}

Magnetic order and excitations in multiferroic DyMnO$_3$ were
studied by neutron scattering experiments using a single crystal
prepared with enriched $^{162}$Dy isotope. The ordering of Mn
moments exhibits pronounced hysteresis arising from the interplay
between Mn and Dy magnetism which possesses a strong impact on the
ferroelectric polarization. The magnon dispersion resembles that
reported for TbMnO$_3$. We identify the excitations at the
magnetic zone center and near the zone boundary in the $b$
direction, which can possess electromagnon character. The lowest
frequency of the zone-center magnons is in good agreement with a
signal in a recent optical measurement so that this mode can be
identified as the electromagnon coupled by the same
Dzyaloshinski-Moriya interaction as the static multiferroic phase.

\end{abstract}

\maketitle

\section{Introduction.}

Strong magnetoelectric coupling may not only imply multiferroic
order, i.e. a phase with coupled magnetic and ferroelectric order,
but it can also result in hybridized collective excitations of
combined polar phonon and magnetic character, which are called
electromagnons \cite{smol82,smol83}. Such excitations may interact
with the electromagnetic radiation through the electric and the
magnetic channels opening the path for multichroism effects
\cite{new1,new2,shuva13}. The fact that these effects can be
controlled through the multiferroic state might become of great
technical relevance. By applying moderate electric fields
multiferroic domains can be easily manipulated
\cite{yamasaki06a,hoffmann11,baum14}, allowing one in consequence
to switch the properties of reflected or transmitted
electromagnetic radiation.

\begin{figure*}[t]
\includegraphics*[width=0.7\textwidth]{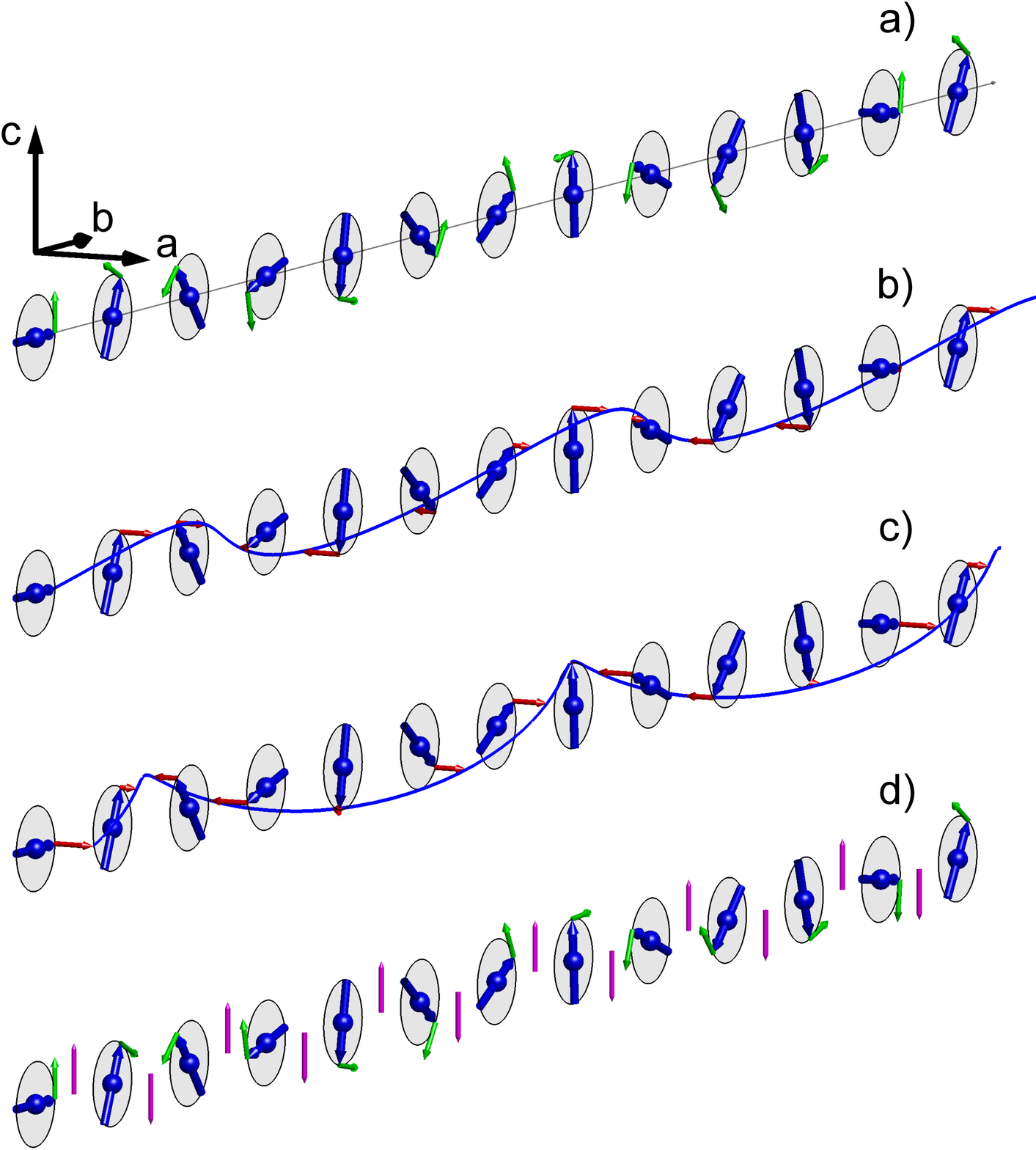}
\caption{(Color online) Illustration of polarization patterns of
several magnon modes which are relevant for the discussion of
electromagnons in DyMnO$_3$. In all parts the static cycloid is
shown by large (blue) arrows rotating in the cycloid plane
indicated by the shaded circles. The propagation vector of the
modulation in $b$ corresponds to that in DyMnO$_3$ with two Mn per
layer and unit cell. Part a) shows the pattern of the phason mode
at the magnetic zone center. The frozen-in oscillating
displacements of the spins are indicated by the small (green
arrows). Adapting the inverse Dzyaloshinski-Moriya interaction
described by equation (1) does not result in an oscillating
polarization, because the cross product
$(\vec{S}_i\times\vec{S}_j)$ is constant. Part (b) shows the CRM
associated with the rotation of the entire cycloid around the $b$
direction. In this mode the spin cross product and thus the
electric polarization rotate around $b$ so that there is an
oscillating polarization along $a$ which can couple with
electromagnetic radiation with $\vec{E}$ parallel $a$. Part (c)
shows the HM in which the spiral plane rotates around the $c$ axis
modifying the character of the spiral from a pure cycloid to a
partial helical one. Applying equation (1) this will yield only a
quadratic coupling with the electric polarization. Part (d) gives
the polarization of the mode polarized within the cycloid plane
($b,c$ plane) appearing at the propagation vector with $k$
component $k_{elm-str}$=1-$k_{inc}$=0.64. This mode gains a strong
oscillating electric polarization through the exchange striction
mechanism. In the static cycloid the scalar product
$\vec{S_i}\cdot\vec{S_j}$ is constant; its momentous deviation
from the average value is indicated by the vertical (magenta)
bars. \label{fig1}}
\end{figure*}

\begin{center}
\begin{figure}[t]
\includegraphics*[width=.90\columnwidth]{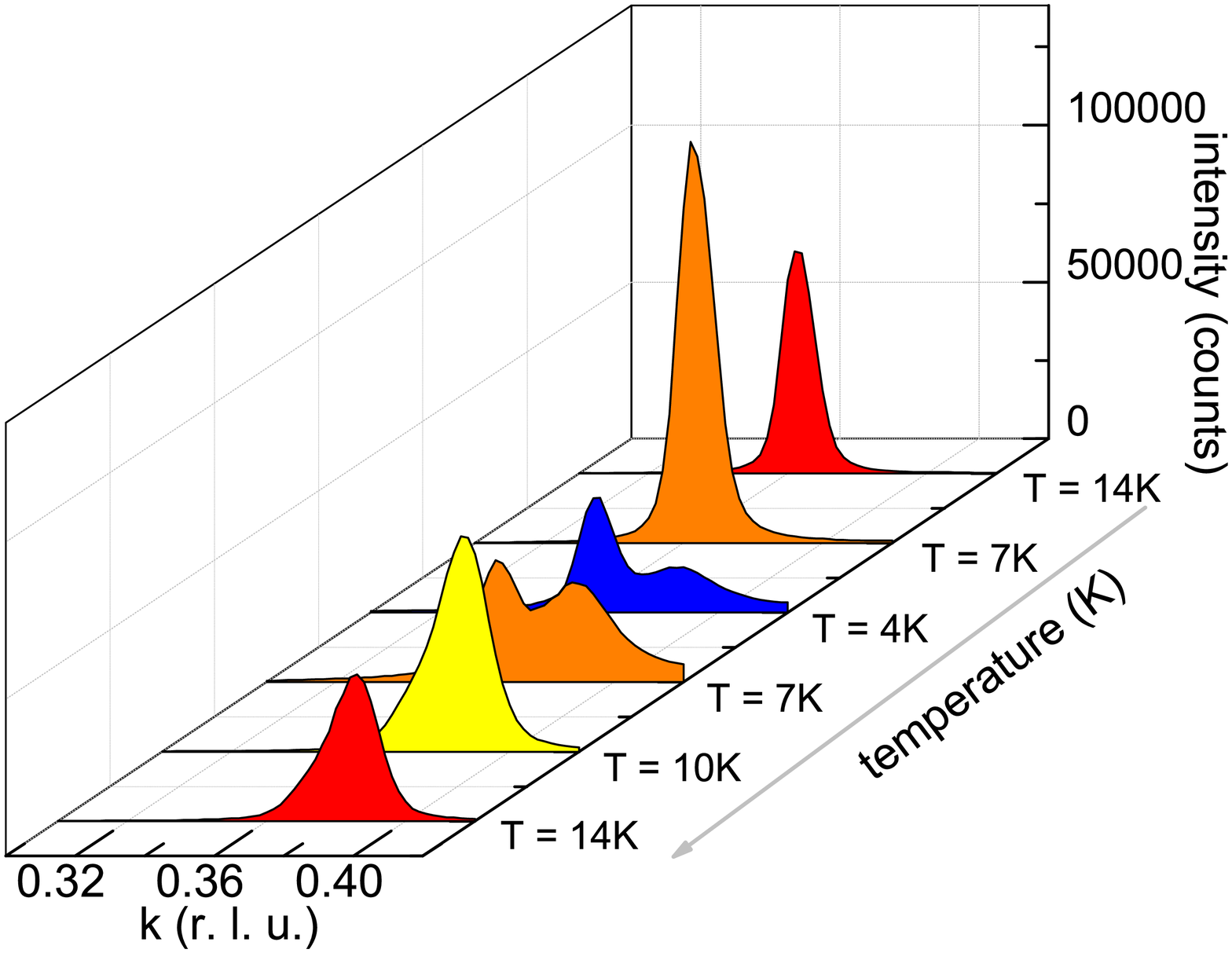}
\caption{(Color online) Temperature dependence of the magnetic (0
k 1) reflection measured from isotopically enriched \dmo \ single
crystal. Data were collected on the 4F2 spectrometer upon cooling
and heating (counting time $\sim$60s), note the different peak
position observed even far above the ordering temperature of Dy
moments. \label{fig2}}
\end{figure}
\end{center}

The understanding of the dynamic magnetoelectric effects, however,
remains too limited. Following the discovery of the multiferroic
phases in  $RE$MnO$_3$ \cite{kimura03a,goto04a,kimura05,cheong07}
with $RE$ a rare earth, first evidence for electromagnon
scattering was reported for this class of materials \cite{pim06}.
It seems intuitive to apply the same magnetoelectric coupling
mechanism, which well explains the static multiferroic phases, to
the collective excitations \cite{katsura07}. In many of the
recently discovered transition-metal-oxide based multiferroics
ferroelectric polarization, $\vec{ P}$, can be explained with the
model of inverse Dzyaloshinski-Moriya (DM) coupling
\cite{katsura05a,mosto06,sergi06} with:

\begin{equation}
 \vec{ P} \propto\vec{r}_{ij}\times(\vec{S}_i\times\vec{S}_j).
 \end{equation}

Here $\vec{ r}_{ij}$ is the distance vector between coupled spins
$\vec{S}_i$ and $\vec{S}_j$. The inverse DM coupling yields only
weak electromagnon scattering due to the small coupling
coefficients arising from spin orbit coupling. Electromagnon
scattering in $RE$MnO$_3$, however, is quite strong and the main
part of it cannot be modified by rotating the static cycloidal
order \cite{val09}. While in \tbmo \ the multiferroic phase is
characterized by a magnetic cycloid in the $bc$ plane in zero
magnetic field \cite{quezel77,blasco00,kenzelmann05a,kajimoto04},
this plane flops to the $ab$ plane \cite{ali09} upon the
application of magnetic fields along the $a$ or the $b$ direction
as well as in the case of other RE's. In contrast the strong
electromagnon response always appears for electric fields along
the $a$ direction \cite{val07,val09}. A large variety of
electromagnon measurements using optical or dielectric techniques
were reported
\cite{pim06,val09,val07,pim06b,kida08,sush07,taka08,lee09,taka09,pim09,taka10,shuva10,rovi10,taka12,taka13}
as well as several theories to explain the strong electromagnon
scattering along $a$
\cite{val09,mochi10,stern09,cano09,stern12,hasega10}. In addition
inelastic neutron scattering gives an insight to the magnetic
character of these modes \cite{senff07,senff08a,senff08b}.

Taking TbMnO$_3$ as the prototypical material for the new class of
multiferroics one may identify at least three electromagnon
contributions in the optical and dielectric
studies\cite{taka08,pim09,shuva10}: at 7.5\ meV there is the
strongest response followed by a signal at 2.7\ meV (at 12\ K). In
addition there is a third electromagnon response at the low energy
of 1.5\ meV \cite{pim09} that is more difficult to detect with
optical methods due to its low energy and due to its small signal
strength. For the contribution at the highest energy Vald\'es
Aguilar et al. \cite{val09} presented a convin\-cing explanation
basing on a modulation of the symmetric exchange, that can be much
stronger than that of the antisymmetric contributions. The
structural distortion associated with the rotation of the \mno
-octhedrons around the $c$ axis generates strong coupling between
a collective displacement of the planar oxygen ions against the Mn
ions and a magnetic mode near the zone boundary of the magnetic
zone \cite{val09}, see discussion below. It is widely accepted
that this exchange-striction mechanism does explain the strong
electromagnon response at high energy. In addition the static
mechanism must also be relevant \cite{katsura07} and explains the
electromagnon signal at the lowest energies
\cite{pim06,pim09,shuva10,taka12,taka13}, which, however,
possesses much lower spectral weight. The intermediate
electromagnon, however, still needs further studies and a signal
at even higher energies is most likely not related to an intrinsic
electromagnon scattering \cite{taka08}.

Comprehensive inelastic neutron scattering experiments in the
cycloidal phases of \tbmo \ yield a microscopic characterization
of the magnon excitations and strong support for the picture
described above \cite{senff07,senff08a,senff08b,kajimoto05a}.
First the modes near the magnetic zone boundary in $b$ direction
indeed exhibit energies where the strongest electromagnon
scattering is found in the optical studies. Three magnon modes can
be identified at the magnetic zone center, e.g. at \vQ = (0 0.28
1) or \vQ = (2 0.28 1), and are illustrated in Fig. 1 a)-c). One
mode can be associated with the oscillation of the phase of the
cycloid, i.e. the phason. This mode cannot induce any
electromagnon character, because the vector product of the
neighboring spins and thus the induced polarization do not vary,
see equation (1) and Fig. 1 a). Two other modes are characterized
by the rotation of the cycloid plane around the $b$ and the $c$
direction, respectively \cite{senff07,katsura07}. Neglecting any
anisotropies the mode for the rotation around the $b$ direction,
see Fig. 1 b), can be considered as the Goldstone mode of the
multiferroic transition and should posses zero energy (being thus
massless) \cite{senff08a}. Applying equation (1) this mode is
associated with the rotation of the ferroelectric polarization
around $b$ which corresponds to an oscillation of the polarization
in $a$ direction. This mode thus exhibits electromagnon character
\cite{senff07,katsura07} and can be measured with optical
techniques for $\vec{E}(\omega)\parallel\vec{a}$. By analyzing the
possible anisotropy energies Senff et al. \cite{senff08a} deduced
that this rotational mode of the cycloid should be the zone center
magnon with the lower but still finite energy, and indeed the few
spectroscopic studies capable to examine the dielectric response
at low enough energies do find excitations at the corresponding
energies \cite{pim06,pim09,shuva10,taka12,taka13}. The other $a$
polarized zone center magnon changes the character of the cycloid
towards a partial helical character, see Fig. 1 c). In order to
differentiate between these two modes we will label the first one
cycloid rotation mode (CRM, see Fig. 1 b)) and the second helical
mode (HM, see Fig. 1 c)). The CRM corresponds to the proposal in
reference \cite{katsura07}. It appears tempting to compare the
intermediate electromagnon scattering in TbMnO$_3$ with the HM as
indeed the frequencies match well. However, the strength of the
dielectric signal does not agree with the character of this mode.
The electromagnon weight could arise through mixing with other
modes \cite{mochi10,val09} but it seems more likely that other
magnon modes cause this intermediate electromagnon response
through the more effective exchange striction mechanism
\cite{stern09,stern12,mochi10}. Due to the flat dispersion between
(0 0 1) and the magnetic Bragg point many magnon modes posses
energies in this range.

\begin{center}
\begin{figure}[t]
\includegraphics*[width=.90\columnwidth]{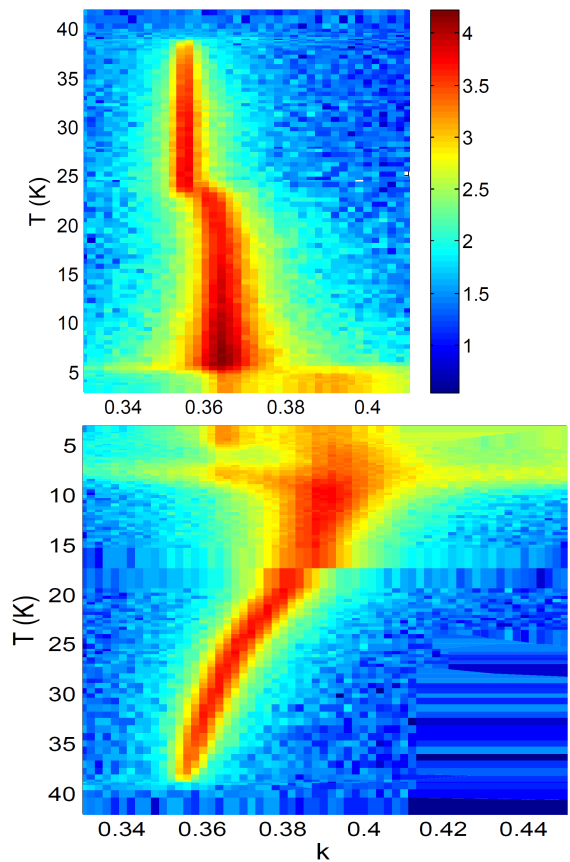}
\caption{(Color online) Temperature dependence of elastic magnetic
scattering in \dmo \ at (0 $k$ 1) studied on the 1T spectrometer
upon cooling (above) and heating (below). At fixed temperatures
the scattering intensity was recorded by scanning along (0 $k$ 1).
The color coding corresponds to a logarithmic scale for a counting
time of about 15\ s; note that the lower scans extend to larger
$k$ values. \label{fig3}}
\end{figure}
\end{center}

While the electromagnon response in the various $RE$MnO$_3$
compounds has been studied with different dielectric or optical
techniques, so far only multiferroic \tbmo \ has been studied by
inelastic neutron scattering, because of the high neutron
absorption of natural Sm, Eu, Gd and Dy. In this work we have used
Dy with enriched $^{162}$Dy isotope content to grow a large
crystal that is suitable for inelastic experiments. The sequence
of magnetic transitions in \dmo \ is similar to that in \tbmo
\cite{quezel77,blasco00,kenzelmann05a,kajimoto04}, with a first
transition to a longitudinal spin-density wave along $b$ at 38\ K
followed by the development of cycloidal order in the $bc$ plane
at 18\ K and the onset of ordering of Dy moments below 10\ K
\cite{feyer06,prokh07,strem07}. However, the interplay between RE
and Mn magnetism seems to be stronger in the Dy compound which is
assumed to contribute to enhanced ferroelectric polarization
\cite{prokh07}. Indeed we find pronounced hysteresis effects in
the Dy as well as in the Mn ordering as function of temperature.
The magnetic excitations and their dispersion also strongly
resemble the findings in \tbmo , which somehow contrasts with the
different electromagnon response in \dmo \ reported by the
dielectric studies.

\section{Experimental}

A large single crystal of \dmo \ was grown by the travelling
floating-zone technique in an image furnace. The feed rod was
synthesized  with a $^{162}$Dy isotope content enriched to 90\%
while a seed crystal with natural Dy was used. The crystal was
characterized by SQUID measurements which may identify the
ordering of Dy moments at 5.4(3) and 7.1(3)\ K upon cooling and
heating, respectively \cite{cronert}. Measurements of specific
heat allow one to determine three ordering temperatures by the
maxima in $c_p /T$ at 5.7(3), 18.8(3) and 38.2(3)\ K
\cite{cronert} in perfect agreement with previously reported
measurements \cite{feyer06,prokh07,strem07} and the neutron
diffraction experiments described below.

Elastic and inelastic neutron scattering experiments were
performed with the cold triple-axis spectrometer 4F2 (\kf =1.55
\Angrz ) and with the thermal triple-axis spectrometer 1T (\kf
=2.66 \Angrz ) at the Orph\'ee reactor in Saclay. At 4F2 a double
monochromator and an analyzer using the (002) reflection of
pyrolithic graphite (PG) were utilized, and a Be filter was set
between the sample and the analyzer in order to suppress
higher-order contaminations. On 1T a PG monochromator, a PG
analyzer and a PG filter were used. Two cylindrical crystals of
about 4\ mm thickness and $\sim$5\ mm diameter were coaligned into
the [010]/[001] scattering geometry for the 4F2 experiment; for
the structural studies on 1T a smaller single piece of the same
growth was mounted. Throughout the paper we use reduced lattice
units to address scattering vectors referring to the orthorhombic
lattice in the setting of $Pbnm$ with $a$= 5.27, $b$=5.83 and
$c$=7.35\AA . The sample was cooled with an orange-type
liquid-helium cryostat and a closed-cycle refrigerator on 4F2 and
1T, respectively.

\section{Results and discussion}

\subsection{Magnetic ordering in \dmo}

The sequence of magnetic transitions in \dmo \ has been studied
using resonant and non-resonant X-ray techniques
\cite{feyer06,prokh07,strem07}. These studies only give access to
the intrinsic ordering of Dy moments and to the structural
distortions accompanying the ordering of Mn moments. In contrast,
the ordering of Mn moments was not studied so far. Compared to
\tbmo \ the emergence of magnetic order is qualitatively similar
but the Dy moments seem to possess a stronger impact even far
above the onset of full order of these moments. Upon cooling,
ordering of Mn moments sets in at \tn =38.2\ K with an
incommensurate modulation along the $b$ direction described by the
propagation vector (0 0.36 0). We study the magnetic ordering near
\vQ =(0 0.36 1) which takes into account the antiferromagnetic
coupling along the $c$ direction. The temperature dependence of
the wavelength of the magnetic modulation was previously studied
via the coupled lattice modulation of half the wave length
corresponding to the doubled pitch of the propagation vector.
Strempfer et al. find a pronounced hysteresis between cooling and
heating cycles \cite{strem07}. With our neutron diffraction
experiment we may directly analyze the ordering of the Mn moments
and we can confirm the strong difference between cooling and
heating cycles, see Fig. 2 , which displays data obtained with the
large sample on 4F2. In the first cooling from room temperature we
find a peak centered near (0 0.363 1) which strongly increases in
intensity upon cooling to 7\ K. Upon further cooling below the
onset of magnetic order of Dy moments, which corresponds to a
different propagation vector of (0 0.5 0), the signal of the Mn
moments seems to split with an additional contribution appearing
at (0 0.39 1). Reheating to 7K, however, does not yield the
initial peak but the additional contribution becomes enhanced and
finally dominant at 10\ K. Even heating to 14\ K does not yield
the initial arrangement although this temperature is well above
the suppression of Dy moment order, which disappears upon heating
at 7.1\ K, as observed in the SQUID data. In order to further
elucidate this complex hysteresis, two full cooling and heating
cycles were recorded on the 1T spectrometer which yielded
identical results, but only one cycle is shown in Fig. 3. The
scans across the signals arising from the Mn moments were fitted
by one or two Lorentzian peaks yielding the results resumed in
Fig. 4.

\begin{center}
\begin{figure}[t]
\includegraphics*[width=.75\columnwidth]{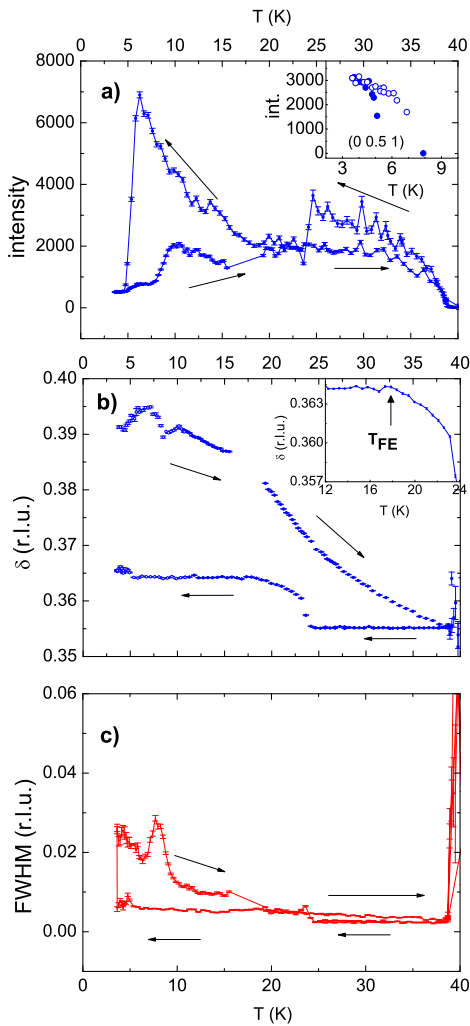}
\caption{(Color online) Temperature dependence of elastic magnetic
(0 $k$ 1) scattering in \dmo \ studied on the 1T spectrometer upon
cooling and heating. The (0 $k$ 1) scans shown in Fig. 3 were
fitted by one or two Lorentzian peak profiles (see Fig. 3). Panel
a) shows the peak intensity (height) of the strongest peak as
function of temperature, in addition the inset displays the height
of the (0 0.5 1) peak that is associated with the ordering of Dy
moments upon cooling and heating with closed and open symbols,
respectively. Panel b) gives the incommensurate pitch (peak
position) of the strongest signal; at low temperatures (open
symbols) the measured profiles were fitted with two Lorentzians.
The inset displays an enlargement of the temperature range near
18\ K corresponding to the onset of ferroelectric order. Panel c)
shows the temperature dependence of the full width at half maximum
(FWHM) of the fitted peak profiles, which senses the impact of the
Dy ordering. \label{fig4}}
\end{figure}
\end{center}

Fig. 3 demonstrates the pronounced hysteresis of the Mn spin order
between cooling and heating which can be perfectly explained by
the coupling of the Dy and Mn magnetism and by a strong coupling
of the anisotropic Dy ions with the lattice. Upon cooling we find
the propagation vector near (0 0.36 1) and an onset of Bragg
scattering at \tn = 38.2\ K, see Fig. 4. Cooling till about 25\ K
results in little variation of the modulation but in a strong
intensity increase, which however suddenly drops at further
cooling accompanied with a step in the modulation vector.  The
second magnetic transition associated with the onset of
ferroelectric order is visible as a kink in the temperature
dependence of the magnetic modulation, see inset of Fig. 4 b),
which stays constant at lower temperatures. The intensity of the
magnetic signal at (0 0.363 1) sharply increases upon further
cooling indicating a stronger modulation of Dy moments with the
same period as that of the Mn moments. This additional intensity
is suppressed at the magnetic ordering of Dy moments near 5\ K,
see inset of Fig. 4a), because Dy moments order at a distinct
propagation vector of (0 0.5 0). The onset of the Dy moment order
is also visible in the broad diffuse scattering accompanying the
transition that extends even beyond the wave vector of the Mn
moments. Concomitantly, the Mn order becomes unpinned and exhibits
a double signal. The additional Mn moment modulation appears at (0
$\sim$0.39 1). This modulation seems to be responsible for the
signal observed in resonant x-ray diffraction at the Dy edge near
(0 2.9 3) which is just the combination of the intrinsic Dy order
and the latter modulation of the Mn moments. The two values
observed for the modulation of Mn moments most likely arise from
strong and minor contribution of Dy moments to the two
modulations, respectively. These results qualitatively agree with
the modulations found at the doubled wave vectors with x-ray
techniques \cite{feyer06,prokh07,strem07}, but the isotope
enriched crystal seems to exhibit slightly smaller
incommensurabilities.

The lower part of Fig. 3 shows the following heating sequence. In
the temperature range till 7\ K there is some variation of the
pitch of the additional Mn signal which becomes dominating upon
heating. Again the loss of Dy order near 7.1\ K, see inset of Fig.
4a), is visible in the broad intensity distribution. Above this
transition, only the Mn signal at larger $k$ remains and it
exhibits a similar intensity compared to that of the initial
signal upon first cooling indicating again a strong contribution
of the Dy moments. But note that the modulation vector of the Mn
order at 10\ K is different for cooling and heating, which
explains hysteretic effects in the ferroelectric polarization
\cite{goto04a}.

The hysteresis of the Mn-moment magnetism in this broad
temperature range is remarkable. It arises from two effects: the
intrinsic hysteresis of the Dy order and the hysteresis of the
coupling of Dy and Mn moments with differing preferred propagation
vectors. The latter results in the two $k$-components of the
propagation vectors of the Mn ordering of 0.36 and 0.39. Upon
cooling the Dy order cannot induce the transformation of the Mn
ordering to the larger values as it happens at too low
temperature. Upon heating the Dy order shifts to 7.1\ K and this
extended temperature range seems sufficient to fully transform the
Mn order to the larger $k$-component. In order to recover the
initial scheme of Mn moment ordering one has to heat up to close
to \tn . The modulation of the Mn order only smoothly decreases
with heating towards the initial values due to the continuous
depinning and due to the melting of Dy moments. Fig. 4c) shows the
temperature dependence of the widths of the fitted peak profiles
at (0 $\sim$0.4 1) arising from the ordering of Mn moments. Above
\tn \ the magnetic scattering of course becomes very broad, but
also near the onset and the disappearance of the Dy order the Mn
magnetism is heavily perturbed. In addition the interplay seems to
imply a finite width of the Mn signal in the almost entire
temperature range, just upon the first cooling slightly below \tn
\ sharp magnetic peaks exist.

Resuming these elastic studies, we find a close coupling between
the magnetism of Dy and Mn moments with very pronounced pinning
effects. These hysteresis and pinning effects posses a clear
impact on the ferroelectric polarization. The pyrocurrent
measurements by Goto et al. \cite{goto04a} were taken upon
heating; they show a huge difference between two runs recorded
after cooling down to 2 and 7\ K, respectively, which reflects the
essential differences in the Mn ordering observed upon cooling and
heating shown in Fig. 2 and 3. Apparently the longer wave-length
magnetic modulation appearing upon cooling results in an up to
$\sim$50\% larger ferroelectric polarization. A strong hysteresis
is also observed in the dielectric constant along $c$ measured on
a thin film of DyMnO$_3$ \cite{lu2013}. In this dielectric
measurement cooling and heating results fall together only above
the N\'eel temperature, resembling our findings for the modulation
vector of Mn moments.

\begin{center}

\begin{figure}[t]
\includegraphics*[width=.90\columnwidth]{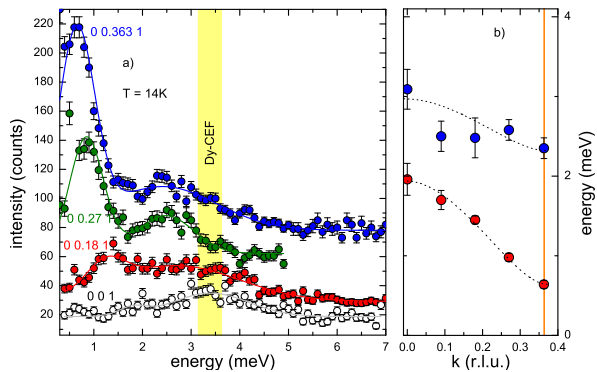}
\caption{ (Color online) Panel a) shows constant \vQ \ scans to
determine the low-energy magnetic excitations between \vQ =(0
0.363 1) and (0 0 1) and dispersion. Slightly above the energy of
3 meV some signal may arise from a Dy crystal field excitation of
the lifting of Kramers doublet degeneracy. Panel b) shows the flat
dispersion in this part of the Brillouin zone. \label{fig5}}
\end{figure}
\end{center}

\subsection{Magnetic excitations in \dmo }

Magnetic excitations were studied on the 4F2 spectrometer using
cold neutrons, $k$=1.55\AA$^{-1}$. The main part of these
measurements were performed at T=14\ K in the multiferroic phase
\cite{kimura05,cronert} but well above the ordering of Dy moments.
Nevertheless Dy moments are already sizeably polarized at this
temperature which implies a more complex magnon dispersion. Data
were taken after cooling the sample to 10\ K only. The
crystal-field excitations in DyMnO$_3$ were not yet studied by
inelastic neutron scattering, and an infrared experiment detected
only one crystal field excitation at 22.9\ meV, which lies far
above the energy range studied here \cite{jandl}. Lower crystal
field energies were just calculated to amount to 5.7 and 15.7\ meV
\cite{jandl}. At the lower of these values we do not find evidence
for a strong crystal field scattering. However, the same
experiment reports a splitting of the Kramers doublet degeneracy
by the interaction between Mn and Dy moments of 3.7\ meV
\cite{jandl}, which might agree with a weak $\vec{Q}$ independent
signal in our scans, see Fig. 5, but for an unambiguous
interpretation further studies are needed. The dispersive features
discussed in the following cannot be attributed to Dy
crystal-field excitations.

In view of the discussion of electromagnons, magnetic excitations
at the zone center, i.e. taken at the magnetic Bragg points, are
most important, as these modes may couple with infinite wavelength
structural distortions even in the harmonic case. Data taken at
\vQ = (0 0.363 1) and other scattering vectors towards \vQ = (0 0
1) are shown in Fig. 5. At the magnetic zone center one may
recognize a low-energy mode at 0.8\ meV and another one near 2.3\
meV. At this scattering vector modes polarized in $a$ direction
fully contribute while a mode polarized in the $c$ direction is
strongly suppressed. Note that in general neutron scattering only
senses the magnetic components polarized perpendicular to the
scattering vector \cite{marshall71}. It appears thus reasonable to
identify the two observed modes with the $a$ polarized modes
arising from the rotations of the cycloid planes \cite{senff07},
CRM and HM, see Fig. 1 b) and c). Furthermore, the observed
frequencies agree with those reported in \tbmo \ at slightly
higher temperature: energy of HM at 2.5\ meV and energy of CRM at
1.1\ meV at T=17\ K in \tbmo . Also the rather flat dispersion
between the magnetic zone center and (0 0 1), which can be
discerned for at least two branches, resembles the behavior in
\tbmo . The phason mode, which contributes little to the data in
Fig. 5, must posses a much smaller energy at the magnetic zone
center. Both modes visible in Fig. 5 exhibit larger widths than
what can be explained by the resolution (FWHM $\sim$0.2\ meV).
This can be a consequence of the anharmonic folding discussed in
reference \cite{mochi10}.

\begin{center}
\begin{figure}[t]
\includegraphics*[width=.999\columnwidth]{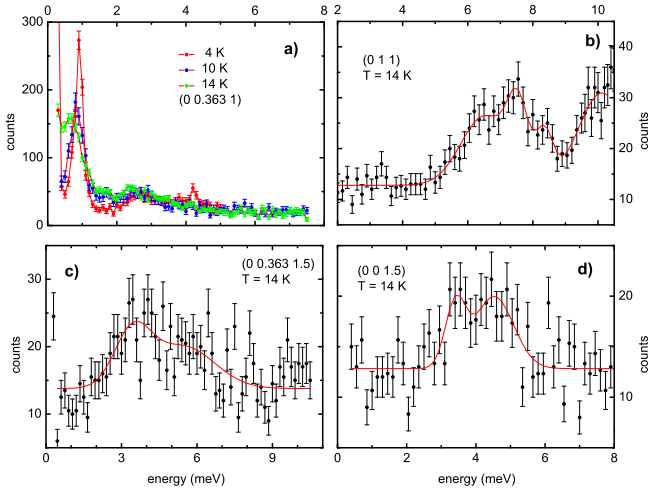}
\caption{(Color online) Constant \vQ \ scans taken at various
point in the Brillouin zone to establish the magnon dispersion in
\dmo ; a) magnetic zone center; b) end point of the dispersion in
$b$ direction;  zone boundary in $c$ direction with c) and without
d) the incommensurate modulation along $b$. \label{fig6}}
\end{figure}
\end{center}

Due to the less favorable scattering conditions with the small
sample only the dispersion along the $b$ direction and the zone
boundary along $c$ could be studied. The scans taken at the latter
positions, (0 0.363 1.5) and (0 0 1.5) indicate a double signal
near 4\ meV which again agrees with the dispersion reported for
\tbmo . In this direction only the antiferromagnetic coupling
along the $c$ direction intervenes, which amounts to about
$J_{AFM}=-0.5$\ meV and which is little reduced between LaMnO$_3$
\cite{moussa96a} and \tbmo \cite{kajimoto05a,senff08a} , as well
as between \tbmo \ and \dmo .

The total dispersion along $b$ ranging from \vQ =(0 0 1) and the
magnetic Bragg to the boundary at \vQ = (0 1 1) is shown in Fig.
7. The upper panel displays the raw data and the lower part the
dispersion of the maxima obtained by fitting gaussian intensity
distributions. As for TbMnO$_3$ a split dispersion consisting of
at least two or three branches can be followed across the entire
zone reaching energies of 6 to 8\ meV at the zone boundary, see
Fig. 6b) where the signal at higher energy seems to arise from a
phonon. This suggests that the ferromagnetic interaction between
nearest neighbors in the $ab$ planes, $J_{FM}$, is little
renormalized when passing from the Tb to the Dy compound, in
agreement with the small difference of the ionic radii
\cite{senff08a}. The main difference in the magnon dispersion
arises from the enhanced frustration by the next-nearest neighbor
interaction along $b$, $J_{NNN}=-\eta \cdot J_{FM}$, which results
in a larger value of the incommensurate modulation vector in \dmo
, $q_k$=0.36 instead of 0.28 in \tbmo . In DyMnO$_3$ the
interaction ratio $\eta$ amounts to 1.15 in comparison to the
value of 0.78 in TbMnO$_3$. This enhanced frustration is in full
agreement with the stronger structural distortion arising from the
slightly smaller ionic radius in \dmo . Further parameters needed
to describe the dispersion are the antiferromagnetic exchange in
$c$ direction, $J_{AFM}$, the single-ion anisotropy term
$\Lambda$, and the spin value S=2 for Mn$^{3+}$.  The simple
spin-only Hamiltonian of the magnetic interaction is given by:

\begin{equation}\label{eq-hamiltonian-moussa}
    \mathcal{H} = -\sum_{i,j}J_{i,j}\vec{S}_i\cdot\vec{S}_j -
    \Lambda\sum_i{S_i^z}^2,
\end{equation}
where the sum only contains the nearest neighbors in the $a,b$
planes, $J_{FM}$, nearest neighbors parallel to $c$, $J_{AFM}$,
and the next-nearest neighbors in the planes $J_{NNN}$. Besides
for $\eta$, only slight modifications are needed to describe the
magnon dispersion in TbMnO$_3$ and \dmo , see Fig. 7.

\begin{center}
\begin{figure}[t]
\includegraphics*[width=.75\columnwidth]{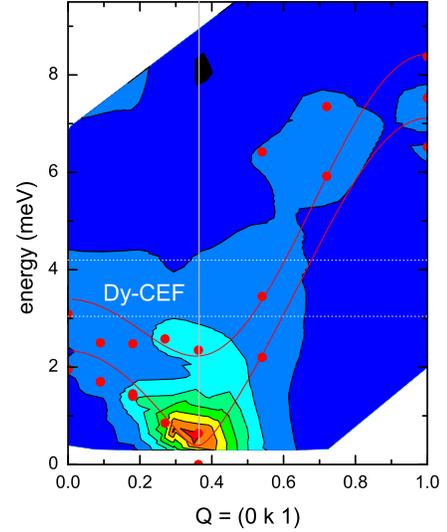}
\caption{(Color online) Dispersion of magnetic excitations along
the $b$ direction shown as a color contour plot of the scattered
intensities. The lines denote the spin-wave dispersion calculated
in linear theory as described in reference \cite{senff08a}.
Parameters were $J_{FM}$=0.12(0.14)meV, $J_{AFM}$=-0.4(-0.35)meV
and $\Lambda$=0.19(0.4)meV for the lower(upper) curves; red dots
denote peak positions determined by fitting Gaussians to the
measured scans.\label{fig7}}
\end{figure}
\end{center}

Most interesting are the zone-center magnons as potential
candidates for electromagnon modes. The temperature dependence of
the two $a$ polarized modes taken from the data in Fig. 6a) is
shown in Fig. 8. Both modes harden upon cooling as expected. The
energy of the lowest mode, which can be assigned to the CRM,
perfectly agrees with the low-energy electromagnon signal detected
recently in a THz radiation experiment \cite{shuva13}. As first
proposed in reference \cite{katsura07} the rotation of the entire
cycloid results in an electromagnon mode basing on the same
magnetoelectric DM coupling as the static multiferroic phase, but
the dielectric oscillator strength of this electromagnon mode is
small.

The higher zone center mode near 2.5\ meV, assigned to the HM, can
be compared with electromagnon scattering observed in THz
spectroscopy, see reference \cite{kida08}. However, the largest
peak in the THz radiation experiment (tracing $\epsilon_2$)
clearly exhibits a lower energy than the zone center magnon; for
example at T=14\ K the optical peak appears at $\sim$1.7\ meV well
below the magnon at 2.38(8)\ meV. The HM is not expected to
exhibit strong polarization as the coupling between magnetic
structure and polarization is quadratic. The intermediate-energy
electromagnon scattering constitutes the strongest electromagnon
signal in \dmo \ when tracing $\epsilon_2$ \cite{kida08}, but
considering the oscillator strength both signals are comparable.
In contrast, the high energy signal clearly dominates in \tbmo
\cite{taka08}. Another mechanism than the weak inverse DM
interaction must be responsible for the intermediate optical
signal. It has been proposed that the exchange striction mechanism
combined with some backfolding of magnons is responsible for this
strong low-energy scattering \cite{mochi10}. Indeed the branches
connecting the magnetic zone center to \vQ =(0 0 1) are quite
flat, see Fig. 5b), giving support for such a picture; but this
flat branch lies significantly above the energy of the THz
radiation signal\cite{kida08}.

\begin{center}
\begin{figure}[t]
\includegraphics*[width=.90\columnwidth]{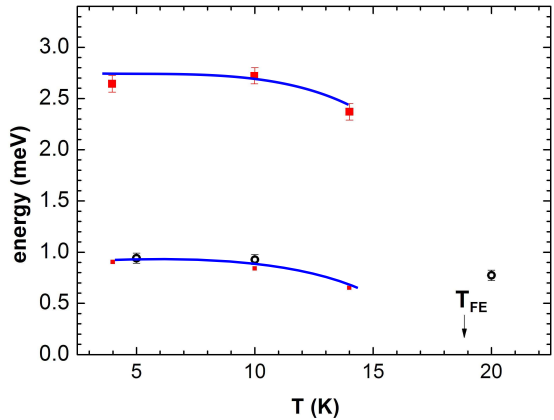}
\caption{(Color online) Temperature dependence of electromagnons
in \dymo : The inelastic neutron scattering results (filled black
symbols) are compared with those of spectroscopic measurements
\cite{shuva13} (open red symbols) for the lowest
 mode. Lines are guides to the eye. \label{fig8}}
\end{figure}
\end{center}

In \dmo , the electromagnon signal found in the spectroscopic
studies at highest energy appears at $\sim$6\ meV at 14\ K, but
this signal only forms a shoulder in $\epsilon_2$ for \dmo .
Following the proposal of Vald\'es Aiguilar et al. this
electromagnon arises from the exchange striction mechanism and a
magnetic arrangement that profits from an alternation of the
magnetic exchange in $b$ direction \cite{val09}. The static
arrangement in a -$\uparrow$-$\uparrow$-$\downarrow$-$\downarrow$-
structure (labelled E-type in $REM$O$_3$ perovskites) couples to
an alternation of enhancement and reduction of the ferromagnetic
interaction in the bonds. This can be seen in the structure
-$\uparrow$-$J_{FM}^{+}$-$\uparrow$-$J_{FM}^{-}$-$\downarrow$-$J_{FM}^{+}$-$\downarrow$-$J_{FM}^{-}$-$\uparrow$-
where parallel moments profit from stronger ferromagnetic
interaction. In a static cycloid with incommensurate modulation
such an oscillating arrangement is not realized for the zone
boundary magnon of the orthorhombic structure but for the magnon
with a $k$ component of the wave vector of
$k_{elm-str}$=1-$k_{inc}$=0.637. This mode is illustrated in Fig.
1 d), which also indicates the variation of the scalar product of
neighboring spin, which alternates. A magnon energy at this $k$
value of the $b$ dispersion indeed amounts to 6\ meV in good
agreement with the spectroscopic data, see Fig. 7. The fact that
there is good agreement at quite different energies in \dmo \ and
\tbmo \ (for Tb this mode appears at 7.5\ meV
\cite{senff08a,taka08}) gives strong support for the exchange
striction mechanism. The difference in the energies in \dmo \ and
\tbmo \ mainly arises from the different incommensurate pitch with
its impact on $k_{elm-str}$. It has been proposed that the strong
low-energy peak in the optical response of \dmo \ and similar
$RE$MnO$_3$ compounds arises from strong backfolding of the magnon
dispersion partially due to the local rotation of the magnetic
anisotropy \cite{mochi10}. The measured dispersion, however, does
not support fully such interpretation as the strong backfolding
effects cannot be observed between (0 0.36 1) and (0 1 1) although
the dispersion is flat between (0 0.36 1) and (0 0 1), see Fig. 5
and 7. Further efforts are needed to fully understand the reasons
of the strong electromagnon response slightly below the frequency
of the HM.

\

\section{Conclusions}

The study of the magnetic ordering confirms the close coupling
between Mn and Dy moments which results in remarkable hysteresis
effects. The intrinsic hysteresis of the Dy moment ordering
together with the depinning of the incommensurate Mn modulation
result in a complex hysteretic behavior which perfectly
corresponds to the reported hysteresis in the temperature
dependence of the ferroelectric and dielectric properties.

The comparison of the inelastic neutron scattering studies on the
magnon dispersion in \dmo \ with previously reported
spectroscopical measurements \cite{kida08,shuva13} helps
identifying the electromagnon modes. The high-energy optical
signal perfectly fits with the energy of the magnon mode that is
expected to strongly couple via the exchange striction mode. Its
energy is significantly lower than that of the corresponding mode
in \tbmo \ mainly due to the enhanced value of the incommensurate
modulation. The zone-boundary magnon energies differ much less
between \dmo \ and \tbmo \ indicating similarly strong
nearest-neighbor ferromagnetic interaction. Concerning the
electromagnon excitation at the lowest energy, there is good
agreement between the dielectric and our neutron scattering
experiments. This mode can be identified as the cycloid rotation
mode which gets its electromagnon weight by the same inverse
Dzyaloshinski-Moriya mechanism as the static multiferroic order.
Again this agrees with the observations for \tbmo \cite{senff08a}.
Concerning the intermediate electromagnon signal, however, there
is no agreement between the spectroscopical studies and the zone
center magnons sensed by neutron scattering. This finding
contrasts with the reports in \tbmo . The strong spectral weight
observed near 2\ meV in the THz radiation studies must possess a
different origin eventually related with back folding of magnon
branches. Indeed the magnon dispersion seems to be quite flat
between (0 $q_{inc}$ 1) and (0 0 1).

\paragraph*{Acknowledgments.} This work was supported by the Deutsche
Forschungsgemeinschaft through Sonderforschungsbereich 608 and
contract AR 613/1-1.


\end{document}